\documentclass[12pt]{report}
\usepackage[dvips]{graphicx}
\newcommand{\sptwo}{1.4}

\newcommand{\doublespace}{\edef\baselinestretch{\sptwo}\Large\normalsize}

\begin{document}
\doublespace
\begin{center}
{\bf On the kinetic  energy of unitary Fermi gas in a harmonic trap
}\\

\renewcommand\thefootnote{\fnsymbol{footnote}}
Alexander L. Zubarev$^1$ and
Michael Zoubarev$^{2}$ \\
 $^1$ Department of Physics, Purdue University\\
West Lafayette, Indiana  47907\\
 $^2$ University of Toronto at Scarborough\\
Toronto, ON, Canada M1C1A4\\
\end{center}
\begin{quote}
We have considered the orbital-free approximation of  the
kinetic energy functional
 to investigate the zero temperature 
properties of dilute harmonically trapped two component Fermi gas at unitarity.
It is shown that our approach provides a realible and inexpensive method to
 study superfluid  strongly interacting dilute
 Fermi gases.
\end{quote}
\vspace{5mm}
\noindent


\pagebreak

The modern density functional theory (DFT) is based on the Kohn-Sham (KS)
 approach [1], where the
noninteracting kinetic energy, $T$, is calculated in terms of the 
KS orbitals, although Ref.[2] 
proved the basic existence of functional $T(\rho)$,
where $\rho$ is the density, $\int \rho(\vec{r}) d^3 r=N$, and $N$ is the
 particle number.
Since  the numerical cost of self-consistently determining
N orbitals rapidly increases for large N,
 the accurate density-functional approximation to the kinetic energy 
 in terms of
the  density  
 would reduce dramatically complexity of the DFT
calculations (here we note the superfluid extension of the DFT given in
Refs.[3-5]). 
There are various noninteracting kinetic energy functionals proposed in
 literature, for chemical 
applications see, for example, Refs.[6-21] (for nonlocal
density functionals see Refs.[22-26]).
For applications of the DFT to the nuclear structure physics see
Ref.[27],
web site, constructed for the universal nuclear energy density functional
 (UNEDF) collaboration, http://unedf.org and references therein.

The kinetic energy functional can be written as
$$
T[\rho]=\frac{\hbar^2}{2 m}\int \tau(\rho(\vec{r}))d^3 r,
\eqno{(1)}
$$
where
the Kirgnitz semiclassical expansion for the kinetic energy  density [28-31]
$$
\tau(\rho(\vec{r}))=
\tau_{TF}(\rho)+\frac{1}{9} \tau_W(\rho)+...,
\eqno{(2)}
$$
where 
$$
\tau_{TF}(\rho)=\frac{3}{5} (3 \pi^2)^{2/3} \rho^{5/3},
\eqno{(3)}
$$
is the well-known Thomas-Fermi (TF) kinetic energy density
and $\tau_W(\rho)$ is the Weizs\"{a}cker kinetic energy density [12]
$$
\tau_W(\rho)=[\vec{\nabla}\rho^{1/2}(\vec{r})]^2.
\eqno{(4)}
$$
The semiclassical expansion
(2)  has to be considered as an asymptotic expansion [11].
While $\tau_{TF}$ is exact for the uniform gas model, $ \tau_W(\rho)$ 
is considered exact in the limit of rapidly varying density $\rho$ [13,14].

We present in this paper the simple linear combinations
$$
\tau=\lambda_1(N) \tau_{TF}+\lambda_2(N)\tau_W
\eqno{(5)}
$$
with $\lambda_{i}$ determined empirically from the asymptotic region and
for getting good energy of few fermion systems at unitarity. A N-dependence
of $\lambda_2$ corresponds to the effective mass N-dependence [32].

There has been a lot of interest in systems of fermions at the unitarity [32-35]
(when the scattering length diverges, the Bertsch
many-body
problem [36]). While Refs.[37-40]
 consider
 homogeneous systems, Refs.[41-45] present {\it ab initio} calculations of the
 properties of trapped fermionic atoms.

In Refs.[46,47] and later in Refs.[48-63] the dynamics of strongly interacting
trapped dilute Fermi gases
 (dilute in the
sense that
 the range of interatomic potential is small
compared with inter-particle spacing) consisting  of a 50-50 mixture of two
 different
 states 
 is investigated in the single equation approach to the time-dependent
density-functional theory, using $\lambda_1=1$ and a constant 
$\lambda_2$ approximations.

For the  stationary case Eq.(5) leads to the following DFT equation
$$
-\lambda_2(N) \frac{ \hbar^2}{2 m} \nabla^2 \Psi
+V_{ext} \Psi+ V_{xc}\Psi=\mu \Psi,
\eqno{(6)}
$$
where $V_{xc}(\vec{r})=[\frac{\partial \rho \epsilon(
\rho)}{\partial \rho}]_{\rho=\rho(\vec{r})}$, $\epsilon(\rho)$ is the
ground-state energy per particle of the homogeneous system,
$
\rho(\vec{r})=\mid \Psi(\vec{r})\mid^2
$
and $\mu$ is the chemical potential.
For the remainder of this paper  we will consider 
fermion systems at unitarity in a 
spherical harmonic trap
$$
V_{ext}(\vec{r})=\frac{m \omega^2 r^2}{2}.
\eqno{(7)}
$$
The ground state energy is given by the minimum of
 the energy functional
$$
J[\Psi]=\lambda_2(N)\frac{ \hbar^2}{2 m}  \int\tau_W(\rho)d^3r+ 
\int V_{ext} \rho d^3r+\int \epsilon(\rho)
\rho d^3r,
\eqno{(8)}
$$
where $\epsilon(\rho)=(\lambda_1(N)+\beta) 3 \hbar^2 k_F^2/(10 m)$,
$k_F=(3 \pi^2 \rho)^{1/3}$, $\rho=|\Psi|^2$, and the universal parameter
$\beta$ [33] is estimated to be $\beta=-0.56$ [39].

To test the accuracy of approximation (5), we write the radial nonlinear equation (6) in  asymptotic region 
$$
(-\lambda_2(N) \frac{ \hbar^2}{2 m} \frac{1}{r^2} \frac{d}{d r} r^2 
\frac{d}{r}+\lambda_2(N) \frac{ \hbar^2}{2 m}\frac{l (l+1)}{r^2}+
\frac{m \omega^2 r^2}{2}-\mu)\Psi=0.
\eqno{(9)}
$$
The regular as $r\rightarrow \infty$ solution of Eq.(9) can be written as 
$z^{-3/4} W_{\kappa, l/2+1/4}(z)$, where $W$ is the regular as 
$r\rightarrow \infty$ Whittaker function, $z=(m \omega/(\hbar \sqrt{\lambda_2(N)})) r^2$ and $\kappa=\mu/(2 \hbar \omega \sqrt{\lambda_2(N)})$,
therefore
$$
\rho \sim e^{-z} z^{2 \kappa-3/2}.
\eqno{(10)}
$$
Since,
in the limit of large $r$ the Hartree-Fock density, $\rho_{HF}(\vec{r})$ is proportional to the square of the last occupied state
$$
\rho_{HF}(\vec{r}) \sim e^{-m \omega r^2/\hbar}r^{2 \mu/(\hbar \omega)-3},
\eqno{(11)}
$$
 we expect that  $\lim_{N\rightarrow \infty}\lambda_2(N) \rightarrow 1$. We 
also expect that
 $\lim_{N\rightarrow \infty}\lambda_1(N) \rightarrow 1$, since the TF 
kinetic energy density becomes exact in the large $N$ limit.
 Note, that in the large $r$ region the density is not slowly 
varying.
For the kinetic energy functional in one spatial dimension Ref.[64] derived an upper bound
$$
<\tau(\rho)>\leq <\tau_W(\rho)>+<\tau_{TF}(\rho)>.
\eqno{(12)}
$$
Although, the question whether the upper bound, Eq.(12), holds in the three 
dimensional case is still open question [65], our numerical results are in 
strong support of the inequality (12), see Fig. 1.

To study the effectiveness of the approximations (5) we calculate lower bounds to the ground state energy, Eg.(8), using the results of Ref.[62]
$$
E^{(-)}=\frac{3}{2} \hbar \omega N \sqrt{\lambda_2(N)+\frac{3 N^{2/3} 
(\lambda_1(N)+\beta)}{4}}.
\eqno{(13)}
$$
To calculate upper bounds, $E^{(+)}$, we employing Fetter's trial functions [66]
$$
\rho^{1/2}(\vec{r})=b (1-(1-q) (d r)^2)^{1/(1-q)},
\eqno{(14)}
$$
where $d$ and $q$ are the variational parameters and $b$ is the normalization constant, to minimize the functional $J$, Eq. (8).

Following Ref.[24], we propose the following approximation for the kinetic 
energy density, $\lambda_2=1$ and
$$
\lambda_1(N)=(1-\frac{1}{N}) (1-\frac{c}{N^{\gamma}}),
\eqno{(15)}
$$
where $c$ and $\gamma$ are fixed by a least squares fit to the fixed-node 
diffusion Monte Carlo data [42],
$c=1.46832$, $\gamma=0.78383$.
It is clear from Table I that the fit is a very accurate. As for the lower
 $E^{(-)}$  and the upper $E^{(+)}$ bounds, they provide the actual solution 
of equation (6), $E=(E^{(+)}+E^{(-)})/2$ within $\pm \delta$ accuracy, with 
$\delta<1\%$. 

Recently, Refs.[58-60] have considered nonlinear equation,  which for stationary case corresponds to the following approximation of the kinetic energy density
$$
\tau=\frac{1}{4} \tau_W+\tau_{TF}.
\eqno{(16)}
$$
Figures 2 and 3 show comparison between energy calculations using approximations (15), (16), (12) and the Thomas-Fermi approximation.
It indicates that (i) there is a very good agreement between calculations using (15) and (16) for $N/2\geq 10$,
(ii) the difference between all four 
approximations is negligible for $N/2 \geq 10^3$. 

The $\tau=(1/4) \tau_W+\tau_{TF}$ approximation has incorrect asymptotic
 behavior for large $r$ and $N$. The success of this approximation (see figures 2 and 3), however,
 indicates
 that the large $r$ behavior have little or no consequence in the calculation 
of ground state energy of N fermion systems at unitarity in a spherical
 harmonic
 trap.

In conclusion, we summarize the main points of this paper.

(i) We have considered the orbital free approximation of the kinetic energy functional, proposed for the first time by Achariya et al. [19], to investigate
N-fermion systems at unitarity
 consisting of 50-50 mixture of two different states and confined in a spherical
harmonic trap. 

(ii) We found that our analytical lower bound, Eq.(13), describes the ground
 state energy with a very good accuracy, providing an easy and simple
 quantitative tool for trapped Fermi gases, without relying on complex and
 extensive computations.

A.L.Z thanks N.J. Giordano and W.L. Fornes for providing the opportunity
 to finish
this work.

\pagebreak

Table I. The energies $E^{(-)}$, Eq. (), $E^{(+)}$,$E=(E^{(-)}+E^{(+)})/2$
 and the energy  calculated within
 the
fixed-node diffusion Monte Carlo method, $E^{MC}$ [42], all in units of
$\hbar \omega$
for $N\leq 30$
[see the text for further details].

\vspace{8pt}

\noindent
\begin{tabular}{lllll}
\hline\hline
$N/2$
&$E^{(-)}$
&$E^{(+)}$
&$E^{MC}$
&$E$  \\ \hline
4
&12.47
&12.51
&12.58
&12.49 \\ \hline
5
&16.69
&16.82
&16.81
&16.76  \\ \hline
6
&21.20
&21.43
&21.28
&21.31 \\ \hline
7
&25.95
&26.29
&25.92
&26.12 \\ \hline
8
&30.92
&31.37
&30.88
&31.15 \\ \hline
9
&36.09
&36.66
&35.97
&36.38 \\ \hline
10
&41.45
&42.14
&41.30
&41.80 \\ \hline
11
&46.98
&47.80
&46.89
&47.39  \\ \hline
12
&52.67
&53.61
&52.62
&53.14 \\ \hline
13
&58.52
&59.58
&58.55
&59.05 \\ \hline
14
&64.51
&65.70
&64.39
&65.10 \\ \hline
15
&70.63
&71.95
&70.93
&71.29
\\ \hline\hline
\end{tabular}

\pagebreak
\begin{figure}[ht]
\includegraphics{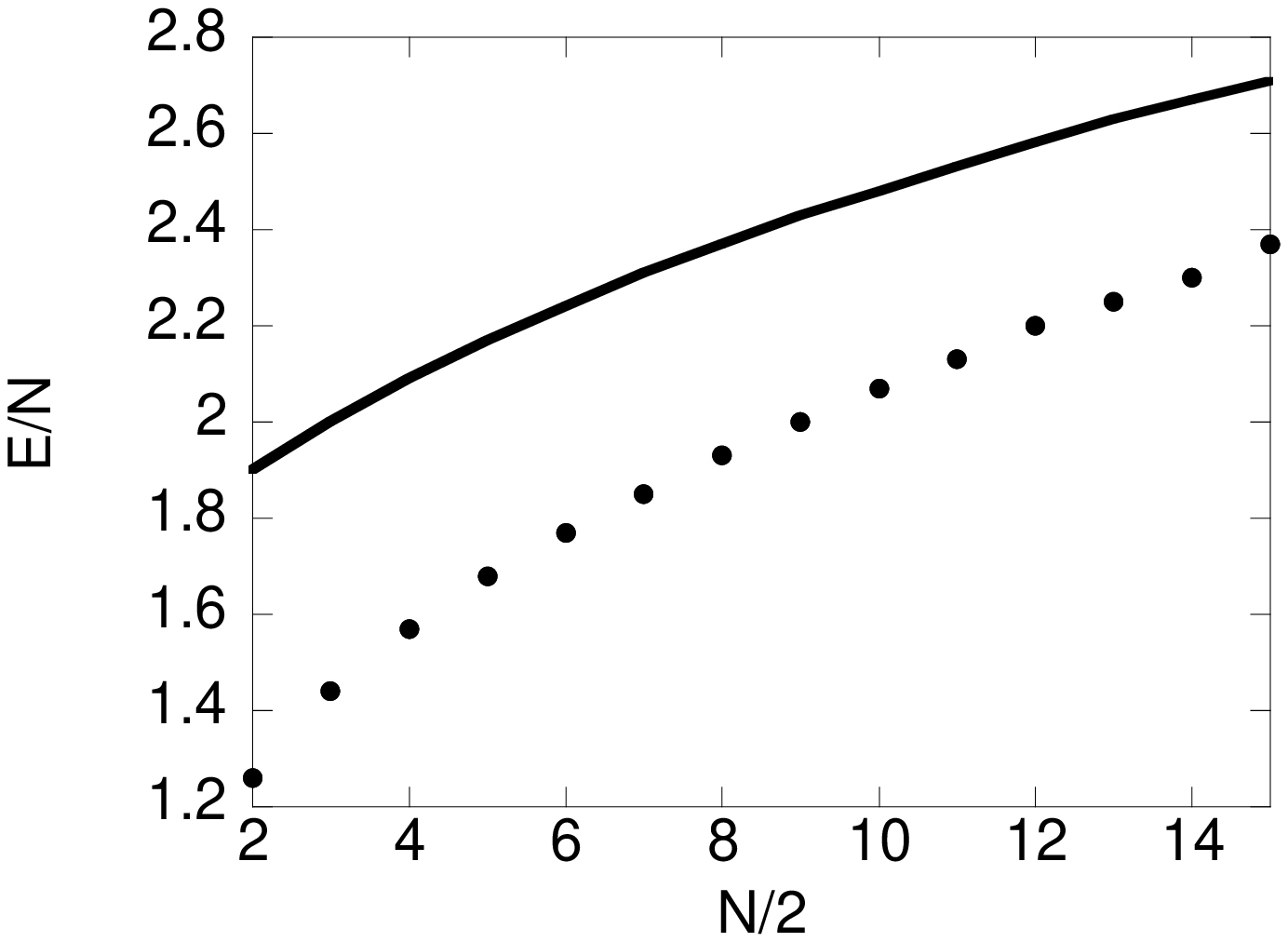}
\end{figure}
Fig.1. Ground state energy per particle of few-fermion systems at unitarity in a
 spherical harmonic trap in units of $\hbar \omega$ as a function of number
of atoms $N$. The solid line represents $\tau=\tau_{TF}+\tau_W $ approximation.
The circular dots indicate
results of Ref.[42].

\pagebreak
\begin{figure}[ht]
\includegraphics{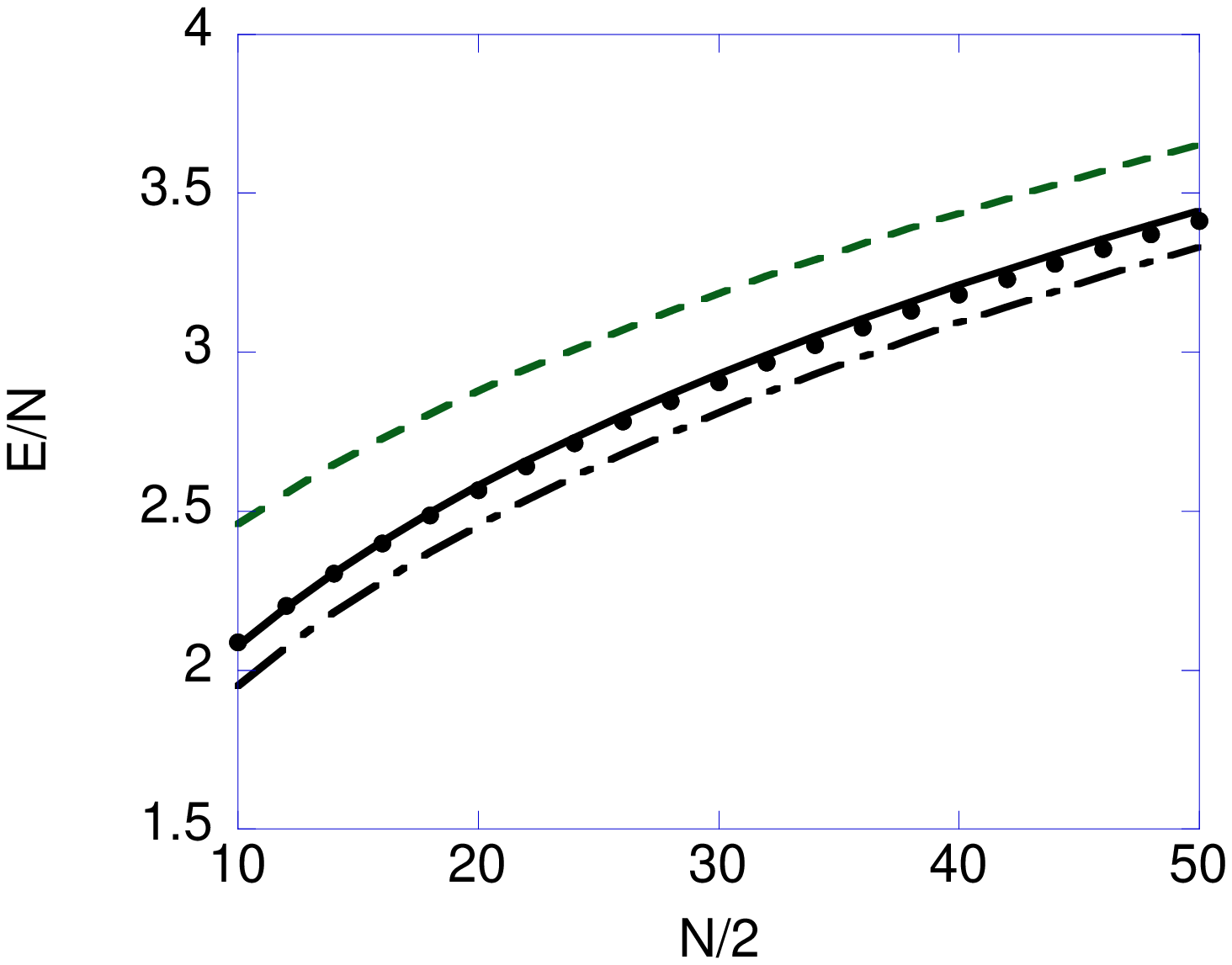}
\end{figure}
Fig.2. Ground state energy per particle 
 in units of $\hbar \omega$ as a function of number
of atoms $N$. The solid line, the circular dots, the dashed line and the 
dashed-dotted line represent results calculated using
 $\tau=\tau_{W}+\lambda_1(N) \tau_{TF}$, $\tau=(1/4)\tau_{W}+\tau_
{TF}$, $\tau=\tau_{W}+\tau_{TF}$ and $\tau=\tau_{TF}$ approximation, respectively.

\pagebreak
\begin{figure}[ht]
\includegraphics{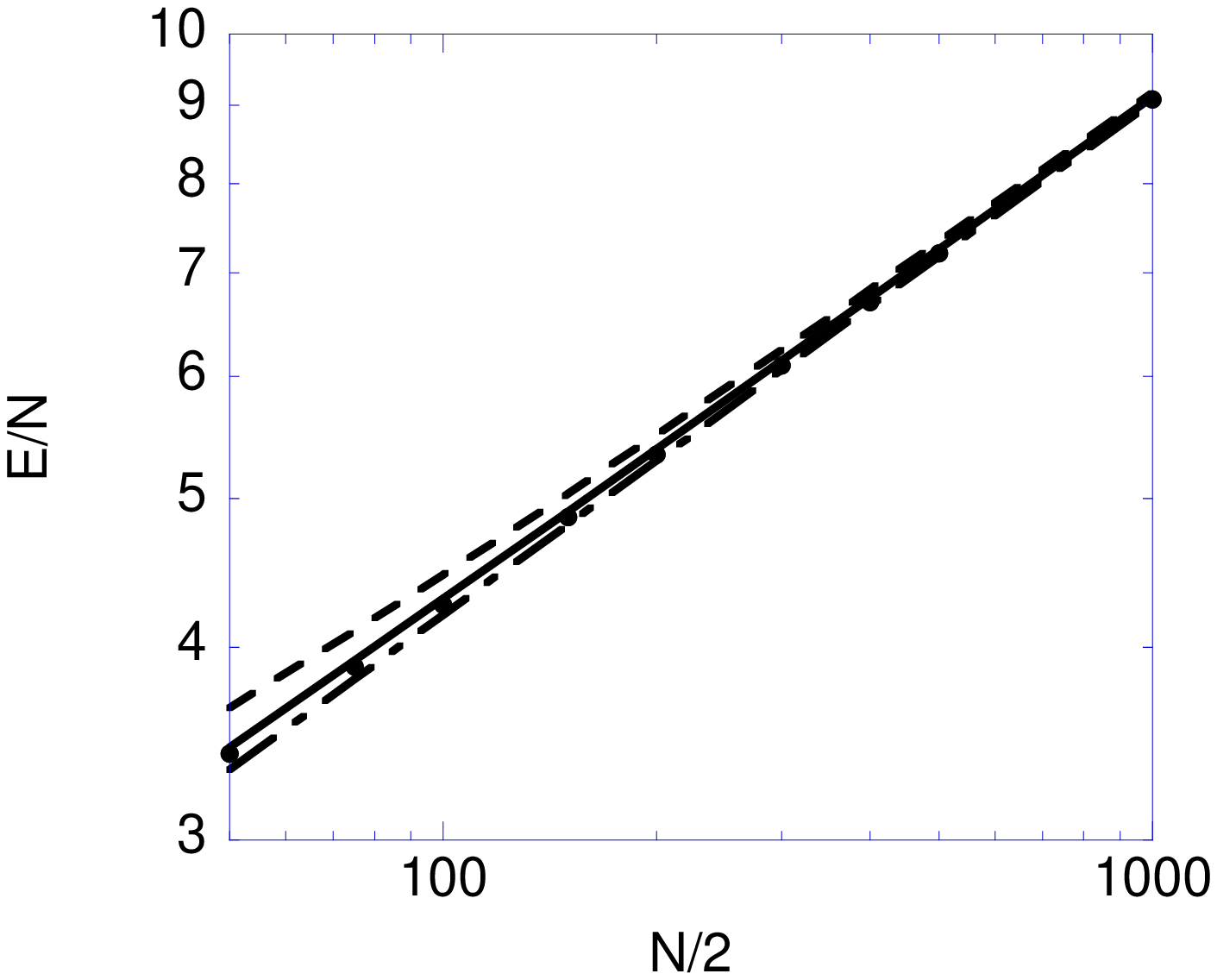}
\end{figure}
Fig.3.  Same as in Fig. 2.

\pagebreak
\noindent
{\bf References}

\vspace{8pt}

\noindent
[1] Kohn W, and Sham L J 1965 {\it Phys. Rev.} {\bf 140}, A1133

\noindent
[2] Hohenberg P, and Kohn W  1964 {\it Phys. Rev.} {\bf 136}, B1133

\noindent
[3] Yu Y, and Bulgac A 2003 {\it Phys. Rev. Lett.} {\bf 90}, 222501

\noindent
[4] Bulgac A  2002 {\it Phys. Rev.} C{\bf 65}, 051305(R)

\noindent
[5] Bulgac A  2007 {\it Phys. Rev.} A{\bf 76}, 040502(R)

\noindent
[6] Kirgnitz D A 1967 {\it Field Theoretical Methods in Many Body Systems}
(London: Pergamon Press)

\noindent
[7] Brack M, and Bhaduri R K 1997 {\it Semiclassical Physics} (Reading MA:
 Addison-Wesley)

\noindent
[8] Brack M, Jennings B K, and Chu P H 1976 {\it Phys. Lett.} B{\bf 65}, 1

\noindent
[9] Hodges C H 1973 {\it Can. J. Phys} {\bf 51}, 1428

\noindent
[10] Guet C, and Brack M  1980 {\it Z. Phys.} A{\bf 297}, 247

\noindent
[11] von Weizs\"{a}cker C F 1935 {\it Z. Phys.} {\bf 96}, 431

\noindent
[12] Parr R G, and Yang W 1989 {\it Density-Functional Theory of Atoms and
Molecules} (New York: Oxford University Press)

\noindent
[13] Dreizler R M, and Gross E K U 1990 {\it Density Functional Theory:
An Approach to the Quantum Many-Body Problem} (Berlin: Springer-Verlag

\noindent
[14]  Yang W.  1986 {\it Phys. Rev.}  A{\bf 34}, 4575 

\noindent
[15] Tomishima Y and Yonei K 1966 {\it J. Phys. Soc. Jpn.}
 {\bf 21}, 142

\noindent
[16] Lieb E H 1981 {\it Rev. Mod. Phys.} {\bf 53}, 603 

\noindent
[17] Chan G K-L, Cohen A J  and  Handy N C 2001 {\it
 J. Chem. Phys.} {\bf
114}, 631 

\noindent
[18] Gazquez J L and  Robles J 1982 {\it J. Chem. Phys.} {\bf 76}, 1467  

\noindent
[19] Acharya  P K et al. 1980 {\it Proc.
Natl. Acad. Sci. U.S.A.} {\bf 77}, 6978

\noindent
[20] Bartolotti L J and  Acharya P K  1982 {\it J. Chem. Phys.} {\bf 77}, 4576

\noindent
[21] Acharya P K 1983 {\it J. Chem. Phys.} {\bf 78}, 2101 

\noindent
[22] Wang L-W and  Teter M P 1992 {\it Phys. Rev.} B {\bf 45} 13196

\noindent
[23] Wang Y A , Govind N , and Carter E A 1998 {\it Phys. Rev.} B
{\bf 58}, 13465; 1999 {\it Phys. Rev.} B {\bf 60}, 16350  

\noindent
[24] Pearson M, Smargiassi E, and Madden P A 1993 {\it J. Physics:
Condens. Matter} {\bf 5}, 3221; Foley M  and  Madden P A 1996 
{\it Phys. Rev.} B {\bf 53}, 10589; Watson S, 
Jesson B J, Carter E A, and  Madden P A 1998 {\it Europhys. Lett.}
{\bf 41}, 37 

\noindent
[25] Choly  N and Kaxiras E 2002 {\it Solid State Commun.} {\bf 121}, 281

\noindent
[26] Chai J-D and Weeks J D 2007 {\it Phys. Rev.} B {\bf 75}, 205122

\noindent
[27] Bertsch G F 2007 {\it J. Phys.: Conf. Ser.} {\bf 78}, 012005 

\noindent

\noindent
[28] Kirgnitz D A 1967 {\it Field Theoretical Methods in Many Body Systems}
(London: Pergamon Press)

\noindent
[29] Brack M, and Bhaduri R K 1997 {\it Semiclassical Physics} (Reading MA:
 Addison-Wesley)

\noindent
[30] Brack M, Jennings B K, and Chu P H 1976 {\it Phys. Lett.} B{\bf 65}, 1

\noindent
[31] Hodges C H 1973 {\it Can. J. Phys} {\bf 51}, 1428

\noindent
[32] Papenbrock T 2005 {\it Phys. Rev.} A{\bf 72}, 041603(R) 

\noindent
[33] O'Hara K M,  Hemmer S L,  Gehm M E,  S R Granade S R, and
Thomas J E 2002 {\it Science} {\bf 298},2179

\noindent
[34] Bartenstein M,  Altmeyer A, Riedl S, Jochim S, Chin C,
Denschlag J H, and Grimm R  2004 {\it Phys. Rev. Lett.} {\bf 92}, 120401

\noindent
[35] Bourdel T, Khaykovich L, Cubizolles J, Zhang J,
Chevy F, Teichmann M, Tarruell L,
Kokkelmans S J J M F, and Salomon C 2004 {\it Phys. Rev. Lett.} {\bf 93}, 050401

\noindent
[36] Bishop R F 2001 {\it  Int. J. Mod. Phys.} {\bf 15}, iii

\noindent
[37] Astrakharchik G E, Boronat J, Casulleras J,
 and Giorgini S  2004 {\it Phys. Rev. Lett.} {\bf 93}, 200404

\noindent
[38] Chang S Y and Pandharipande V R 2005 {\it Phys. Rev. Lett.} {\bf 95},
080402

\noindent
[39] Carlson J, Chang S-Y, Pandharipande V R, and
Schmidt K E 2003 {\it Phys. Rev. Lett.} {\bf 91}, 050401

\noindent
[40] Carlson J  and  Reddy S 2005 {\it Phys. Rev. Lett.} {\bf 95}, 060401

\noindent
[41] Chang S Y, and Bertsch G F 2007 {\it Phys. Rev.} A{\bf 76}, 021603(R)

\noindent
[42] Blume D,  von Stecher J, and Greene C H 2007 {\it Phys. Rev. Lett.}
{\bf 99}, 233201

\noindent
[43] von Stecher J, Greene C H, and  Blume D 2007 {\it Phys. Rev.} A{\bf 76},
053613

\noindent
[44] Blume D  2008 {\it Phys. Rev.} A{\bf 78}, 013635

\noindent
[45] von Stecher J, Greene C H, and  Blume D 2008 {\it Phys. Rev.} A{\bf 77},
043619

\noindent
[46] Kim Y E   and Zubarev A L  2004 {\it Phys. Rev.} A{\bf 70}, 033612

\noindent
[47] Kim Y E   and Zubarev A L  2004 {\it Phys. Lett.} A{\bf 327}, 397

\noindent
[48] Kim Y E   and Zubarev A L  2005 {\it Phys. Rev.} A{\bf 72}, 011603(R)

\noindent
[49] Kim Y E   and Zubarev A L  2005 {\it J. Phys. B: At. Mol. Opt. Phys.}
{\bf 38}, L243

\noindent
[50]  Manini N and Salasnich L 2005 {\it Phys. Rev.} A{\bf 71}, 033625

\noindent
[51] Ghosh  T K and  Machida K 2006 {\it Phys. Rev.} A{\bf 73}, 013613

\noindent
[52] Diana G, Manini N, and Salasnich L 2006 {\it Phys. Rev.} A{\bf 73},
065601

\noindent
[53] Yin J  and Ma Y-L 2006 {\it Phys. Rev.} A{\bf 74}, 013609

\noindent
[54] Salasnich L and Manini N 2007 {\it Laser Phys.} {\bf 17}, 169

\noindent
[55] Zhou Y and Huang G 2007 {\it Phys. Rev.} A{\bf 75}, 023611

\noindent
[56] Ma Y-L  and Huang G 2007 {\it Phys. Rev.} A{\bf 75}, 063629

\noindent
[57] Wen W and  Huang G 2007 {\it Phys. Lett.} A{\bf 362}, 331

\noindent
[58] Wen W,  Zhou Y and Huang G 2008 {\it Phys. Rev.} A{\bf 77}, 033623

\noindent
[59] Adhikari S K  2008 {\it Phys. Rev.} A{\bf 77}, 045602

\noindent
[60] Salasnich L and Toigo F 2008 {\ Phys. Rev.} A{\bf 78}, 053626

\noindent
[61] Adhikari S K and Salasnich L 2008 {\ Phys. Rev.} A{\bf 78}, 043616

\noindent
[62] Zubarev A L 2009 {\it J. Phys. B: At. Mol. Opt. Phys.} {\bf 42}, 011001

\noindent
[63] Wang C, Ma C R and Ma Y 2009 {\it J. Low. Temp. Phys.} {\bf 154}, 85;
2007 { \it J. Phys. B: At. Mol. Opt. Phys.} {\bf 40},  4591 

\noindent
[64] Marsh N H and Young W H 1958 {\it Proc. Phys. Soc. London} A{\bf 72}, 182

\noindent
[65] Lieb E H 1983 {\it Int. J. Quant. Chem.} {\bf 24}, 243

\noindent
[66] Fetter A L 1997 {\it J. Low. Temp. Phys.} {\bf 106}, 643
\end{document}